\documentclass[]{emulateapj}

\begin{document}

\setlength\tabcolsep{7pt}

\title{Tight Correlations Between Massive Galaxy Structural Properties and Dynamics: \\ The Mass Fundamental Plane Was in Place by $z\sim2$}

\author{Rachel Bezanson\altaffilmark{1,2,3}, Pieter G. van Dokkum\altaffilmark{1}, Jesse van de Sande\altaffilmark{4},\\ Marijn Franx\altaffilmark{4}, Joel Leja\altaffilmark{1}, and Mariska Kriek\altaffilmark{5}}

\altaffiltext{1}{Department of Astronomy, Yale University, New Haven, CT 06520-8101, USA}
\altaffiltext{2}{Steward Observatory, University of Arizona, AZ 85721, USA}
\altaffiltext{3}{Hubble Fellow}
\altaffiltext{4}{Sterrewacht Leiden, Leiden University, NL-2300 RA Leiden, The Netherlands}
\altaffiltext{5}{Department of Astronomy, University of California, Berkeley, CA 94720, USA}

\shortauthors{Bezanson et al.}
\shorttitle{The Mass Fundamental Plane Was in Place by $z\sim2$}

\slugcomment{Accepted to ApJ Letters}

\newcommand{\unit}[1]{\ensuremath{\, \mathrm{#1}}}

\begin{abstract} 
The Fundamental Plane (FP) is an empirical relation between the size, surface brightness, and velocity dispersion of early-type galaxies.  This relation has been studied extensively for early-type galaxies in the local universe to constrain galaxy formation mechanisms. The evolution of the zeropoint of this plane has been extended to high redshifts to study the luminosity evolution of massive galaxies, under the assumption of structural homology. In this work, we assess this assumption by replacing surface brightness with stellar mass density and present the evolution of the ``mass FP'' for massive, quiescent galaxies since $z\sim2$. By accounting for stellar populations, we thereby isolate and trace structural and dynamical evolution. Despite the observed dramatic evolution in the sizes and morphologies of massive galaxies since $z\sim3$, we find that quiescent galaxies lie on the mass FP out to $z\sim2$. In contrast with $\sim1.4\unit{dex}$ evolution in the luminosity FP, average residuals from the $z\sim0$ mass FP are less than $\sim0.15\unit{dex}$ since $z\sim2$. Assuming the \citet{hyde:09} mass FP slope, we find that this minimal offset scales as $(1+z)^{-0.095\pm0.043}$. This result lends credence to previous studies that derived luminosity evolution from the FP. Therefore, despite their compact sizes and suggestions that massive galaxies are more disk-like at $z\sim2$, the relationship between their dynamics and structural properties are consistent with local early-type galaxies.  Finally, we find no strong evidence for a tilt of the mass FP relative to the Virial plane, but emphasize the need for full models including selection biases to fully investigate this issue.
\end{abstract}


\section{Introduction}

Strong correlations between the structural properties and dynamics of early-type galaxies have been extensively demonstrated, most notably in the Fundamental Plane (FP), which connects the surface brightness, size, and velocity dispersion of these massive galaxies \citep[e.g.][]{djorgovski:87,dressler:87}. The ``tilt'' and zeropoint of the FP with respect to the expected Virial plane (VP) are related to the M/L ratios of these massive galaxies. Although the form of the FP does not evolve strongly \citep[e.g.][]{holden:10}, the zeropoint of the relation evolves dramatically with redshift.  This evolution has been modeled as surface brightness evolution due to aging stellar populations of early-type galaxies \citep[e.g.][]{dokkum:96,holden:10,toft:12}, although it has also been suggested to be a result of galaxy size evolution \citep{saglia:10}.  

Due to the availability of stellar mass estimates from broadband photometry, surface brightness can be replaced with stellar mass density in the ``mass fundamental plane'' (mass FP) \citep[e.g.,][HB09]{hyde:09}. The mass FP is closer to the Virial Plane (e.g. HB09).  Furthermore, it is not sensitive to aging stellar populations and is stable in galaxy merger simulations; once galaxies lie on the mass FP they are expected to evolve along that relation \citep[e.g.][]{boylan:06,robertson:06a}.  Recent observations have uncovered the dramatic morphological evolution of massive galaxies since $z\sim2$, both in size \citep[see e.g.][and references therein]{dokkumnic:08} and possibly in shapes \citep[see e.g.][]{wel:11,chevance:12}. Such evolutionary effects are not explicitly included in simulations of FP evolution and could move galaxies perpendicular to the FP.

In this paper, we observationally test the stability of the mass FP since $z\sim2$. This work includes a compilation of data from the literature in conjunction with a new sample of galaxies at $z\sim0.7$ with spectroscopic data from the DEIMOS spectrograph on Keck II and HST (ACS/WFC3) imaging. 

We assume concordance cosmology ($H_0=70\rm{\,km/s\,Mpc^{-1}},\Omega_M=0.3\,\&\,\Omega_{\Lambda}=0.7$) and AB magnitudes.

\section{Data}

\subsection{$z\sim0$: SDSS}
The local galaxy sample is drawn from the Sloan Digital Sky Survey DR7 \citep{dr7} and includes galaxies at $0.05<z<0.07$. Galaxy effective radii are $r'$ band S\'ersic fits from \citet{simard:11} and stellar mass-to-light ($M/L$) ratios are taken from the MPA-JHU catalog \citep{brinchmann:04} and scaled to the best-fit Sersic magnitudes \citep[see e.g.,][]{taylor:10}.  Spectroscopic redshifts and velocity dispersions are taken from the Princeton pipeline. All velocity dispersions are aperture-corrected to $R_e$ by: 
\begin{equation}\sigma_{R_e}=\sigma_{ap}(r_{ap}/R_e)^{0.066}\label{eqn:apcor}\end{equation}
as derived by \citet{cappellari:06} for local early-type galaxies in the SAURON sample.  Absolute $g'$-band magnitudes are calculated using NYU-VAGC Kcorrect AB $g'$-band magnitude, K-correction \citep{blanton:07}, the distance modulus, $DM(z)$
\begin{equation}M_{g'}=m_{g'}-DM(z)-K_{g'}(z).\end{equation}

\subsection{Velocity Dispersion-Selected Sample at $z\sim0.7$}\label{sec:deimos}

We observed $\gtrsim100$ galaxies at $0.4<z<0.9$ from the Newfirm Medium Band Survey (NMBS)-Cosmos \citep{whitaker:11} and UKIDSS UDS fields \citep{williams:09}, focusing on overlap with the CANDELS \citep{candels} fields, using DEIMOS on Keck II from January 19-21, 2012. Three masks were observed: two in NMBS-Cosmos with total exposures of 13.67 and 5.67 hours and one in UDS for 7.67 hours using the $1200\unit{mm^{-1}}$ grating, centered at $7800\unit{\AA}$. Galaxies were selected to span a range in \emph{inferred velocity dispersion} \citep[see e.g.][]{bezanson:11}, $\sigma_{inf}\gtrsim100\unit{km\,s^{-1}}$. This corresponds to a selection in size and mass, with no preselection on morphology. Therefore the dataset includes early and late-type and quiescent and star-forming galaxies. Only the 54 quiescent galaxies with $\leq10\%$ errors in velocity dispersion and HST imaging are included in this Letter, the complete dataset will be fully described and published in Bezanson et al. in prep.\footnote{Subsample coordinates and velocity dispersions can be found at http://www.astro.yale.edu/bezanson/Bezanson+13b.html}

Spectra were reduced and extracted using the Spec2d pipeline \citep{cooper:12, newmandeep:12}. Telluric corrections are applied by fitting models for atmospheric absorption, scaled to fit spectra in each mask. The instrumental resolution, as measured from sky lines, was $\sim1.6\unit{\AA}$ at $\sim7800\unit{\AA}$, which corresponds to $R\sim5000$ or $\Delta\,v\sim60\unit{km\,s^{-1}}$ with a spectral range of $\sim6500-9200\unit{\AA}$. Velocity dispersions were measured using PPXF \citep{cappellari:04} to fit \citet{bc:03} SPS models.  Measured velocity dispersions are corrected for the template dispersions and are aperture corrected to $R_e$ using the best-fit S\'ersic profile from CANDELS F160W WFC3 imaging \citep{wel:12} when available (28 galaxies) or ACS F814W imaging \citep{bezanson:11} in NMBS-Cosmos (24 galaxies). Errors on velocity dispersions do not include additional effects such as template mismatch, which will be fully discussed in Bezanson et al. in prep.

\begin{figure*}[t]
  	\includegraphics*[width=\textwidth]{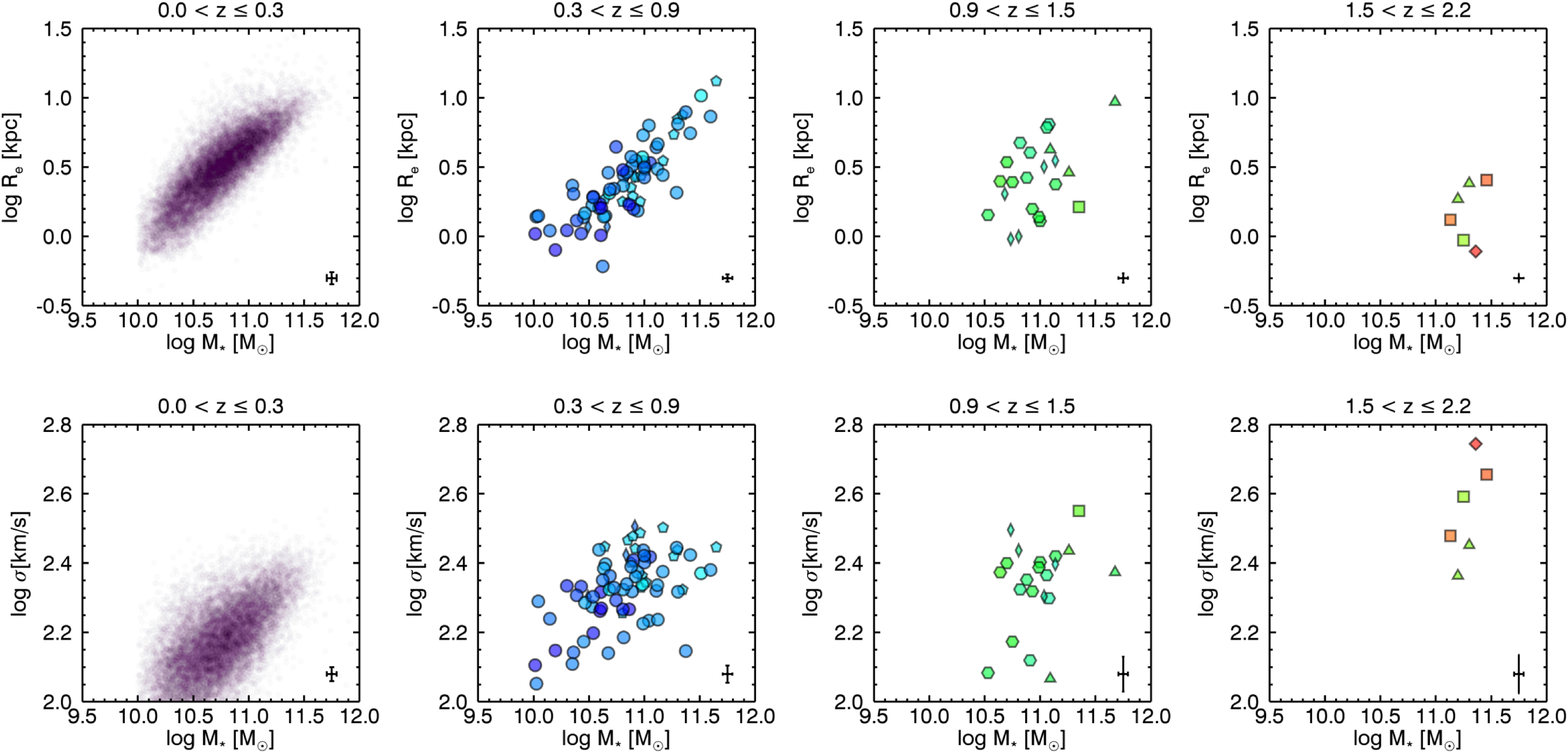}
	\includegraphics*[width=\textwidth]{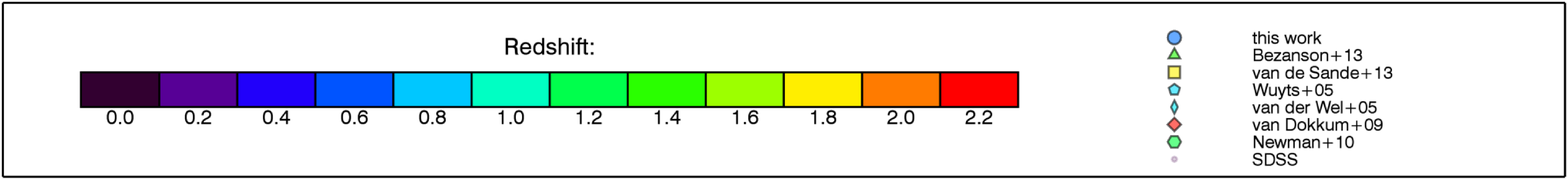}
    \caption{Structural and dynamical measurements included in the mass FP. Symbol colors correspond to galaxy redshift and shapes to data sample. \emph{Top Row:} $\log\,R_e$ versus $\log\,M_{\star}$ and \emph{Bottom Row:} $\log\sigma$ versus $\log\,M_{\star}$ in redshift bins to $z\sim2$.}
  \label{fig:structs}
 \end{figure*}

\subsection{Literature Data at $z>0$}
We compiled a sample of 64 galaxies spanning $0.5<z<2.2$ with spectroscopic velocity dispersion measurements.  We ensured that derived properties, such as aperture corrections to measured velocity dispersions, stellar masses, and rest-frame colors are calculated uniformly.   Best-fit S\'ersic parameters are measured from HST imaging. Spectral energy distribution (SED) fitting for stellar masses and rest-frame colors are calculated uniformly following e.g. \citet{whitaker:11}, with all luminosities corrected to reflect the total luminosity of the best-fit S\'ersic profile.

We match 17 galaxies from the \citet{wuyts:04} sample to the FIRES photometric catalog \citep{forster:06} and S\'ersic fits from \citet{blakeslee:06}, excluding galaxies with ``Merger/Peculiar'' morphologies to eliminate poor fits. \citet{blakeslee:06} report both Galfit and photometric I band (F814W) magnitudes, which we further correct with the ratio of ``auto'' to total flux in the Ks band in the FIRES catalog for each galaxy.

The \citet{wel:05} sample (17 galaxies) is matched to the photometric 3D-HST v.2.1 catalog in GOODS-S (Skelton et al. in prep) and \citet{wel:12} catalog of Sersic fits to CANDELS F160W WFC3 imaging. Luminosities are corrected to the S\'ersic magnitudes using photometric F160W magnitudes. 

The single galaxy in \citet{dokkumnature:09} is in the MUSYC 1255 catalog \citep{blanc:08}.\footnote{1255-0 is at the edge of the MUSYC field and is excluded from the public catalog.} A best-fit Sersic profile is fit to NICMOS2 imaging in the H160 band. We do not have a calibrated S\'ersic flux for this galaxy, nor are there photometric measurements from the NICMOS imaging, however the effective radius of 1255-0 is small ($\sim0.1''$) so we expect the photometric measurement to be excellent from the MUSYC imaging.

The 16 galaxies in the \citet{newman:10} sample are located in the EGS (Extended Growth Strip), GOODS-N, and SSA22 fields. Photometry and morphological measurements (de Vaucouleurs profiles) in the ACS F814W (EGS and GOODS-N) and F850LP (SSA22) bands in this sample are provided in \citet{newman:10} and by private communication with the author.  

The \citet{sande:11,sande:13} sample includes five galaxies from NMBS-Cosmos catalogs \citep{whitaker:11} and UKIDSS UDS catalogs \citep{williams:09}. For four of these galaxies, S\'ersic profiles are fit to F160W WFC3 imaging.  For one galaxy for which HST imaging is unavailable, the size is measured from ground-based UKIDSS K band imaging. S\'ersic luminosity corrections are provided in the original sample. 

Finally, we include eight galaxies at $z\sim1.5$ from \citet{bezanson:13}, also from NMBS and UDS catalogs. For seven of these galaxies, S\'ersic profiles are fit to F160W WFC3 imaging, with aperture magnitudes directly measured from the same imaging.  For one galaxy the size and S\'ersic luminosity are measured from F814W ACS imaging.

\subsection{Derived Properties: Sizes, Stellar Masses, \& Rest-frame colors}\label{s:sizemass}

Galaxy sizes are defined as the circularized effective radius, $R_e=\sqrt{ab}$ from S\'ersic fits.  We note that although traditionally $R_e$ is defined in FP studies by fitting $r^{1/4}$ profiles, fitting the more flexible S\'ersic profiles has been shown to give consistent FP parameters \citep[e.g.][]{kelsonfp:00}.

All published velocity dispersions are aperture corrected to $R_e$, following equation (\ref{eqn:apcor}).

Stellar M/L estimates are calculated for each SED using FAST \citep{kriek:09} and corrected to the best-fit S\'{e}rsic luminosity to account for missing flux in the photometric catalogs.  For all samples, we assume delayed exponential star-formation histories, a \citet{chabrier:03} IMF and use \citet{bc:03} templates. Rest-frame colors are calculated using InterRest \citep{taylor:09}. 

\subsection{Selection of a Quiescent Sample}
For this study of the FP, we only include galaxies that are dominated by quiescent stellar populations, as defined by their rest-frame colors. For SDSS galaxies, we isolate red sequence galaxies using $^{0.1}(g-r)$ color cuts from \citet{graves:07}. We note that using structural cuts at $z\sim0$ instead (following e.g. HB09) does not impact the conclusions of this Letter; changes of derived FP parameters are within quoted uncertainties. In order to distinguish between red galaxies at high-z which are quenched, and those which have significant dust-reddening, we use the SED shape in the near-IR. This method has been shown to effectively isolate quiescent galaxies \citep[e.g.][]{williams:09,whitaker:12a}. 

We adopt U-V and V-J color cuts from \citet{whitaker:12a}, defining quiescent galaxies as $U-V>1.3$, $V-J<1.5$, and $U-V>0.8(V-J)+0.7$. Although these spectroscopic samples primarily consist of quiescent galaxies, this eliminates a fraction of galaxies, particularly for the DEIMOS sample (\S \ref{sec:deimos}).  

The final sample consists of 16598 quiescent galaxies in the SDSS and 102 from $0.3<z<2.19$. Fig.\ref{fig:structs} shows the wide range of properties of galaxies in the combined sample.  We include galaxies with $\log\,M_{\star}\geq10$, however the effective mass limits of spectroscopic surveys increase with redshift as longer exposures are required to obtain sufficient S/N for dynamical measurements.  In the top row of Fig.\ref{fig:structs}, the evolution of galaxy size at fixed stellar mass \citep[see e.g.][]{dokkumnic:08} is prominent for this sample.  The bottom row of Fig.\ref{fig:structs} indicates that galaxies in this sample have somewhat higher velocity dispersions at fixed mass with redshift. We note that apparent evolution in size and velocity dispersion is a combination of selection bias and intrinsic evolution.  Incompleteness in spectroscopic samples can also be influenced by galaxy compactness, and therefore velocity dispersion, which boosts S/N as more light enters the spectrograph aperture, biasing towards more compact galaxies with higher velocity dispersions \citep[see e.g.][for more in-depth analysis of these biases]{sande:13}.

\section{Fundamental Plane in Luminosity and Mass}

 \begin{figure*}[!t]
   \centering
	\includegraphics*[width=\textwidth]{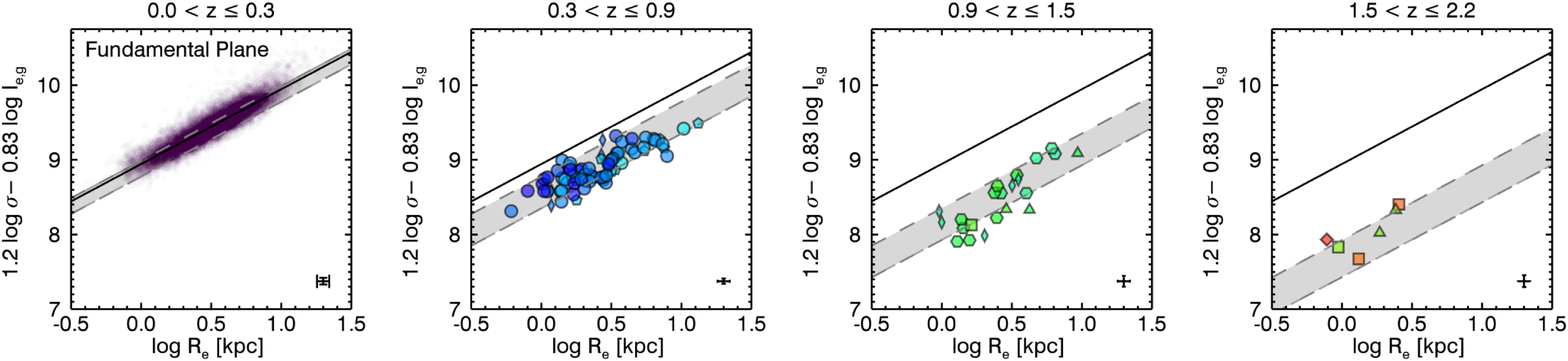}
	\includegraphics*[width=\textwidth]{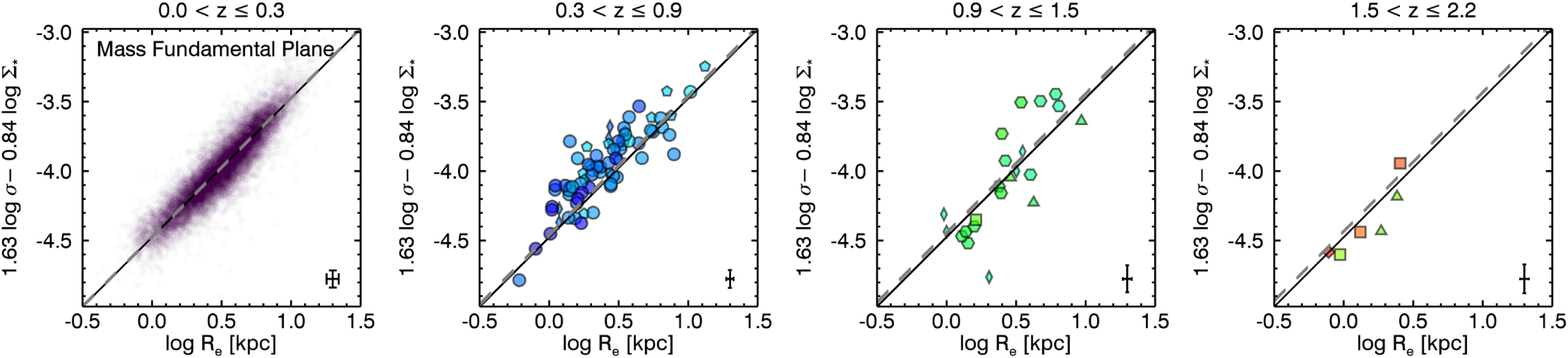}
	\includegraphics*[width=\textwidth]{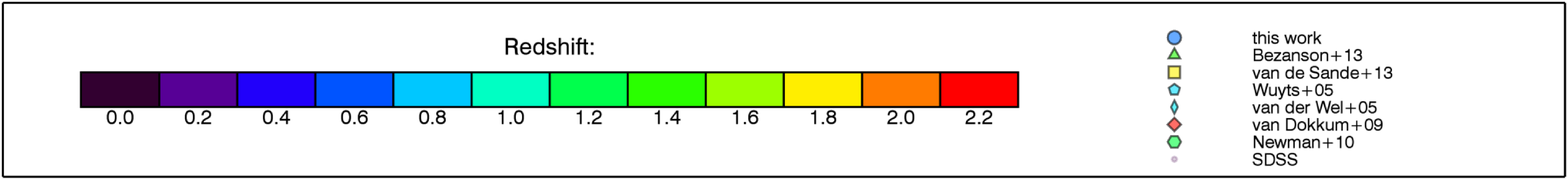}
	\caption{FP and mass FP. Symbols as in Fig.\ref{fig:structs}. Lines indicate $z\sim0$(solid black) and redshift-dependent (gray dashed and shaded regions) relations. \emph{Top Row:} Rest-frame $g'$-band FP. Assuming no evolution in slope, the FP zeropoint evolves strongly with redshift. \emph{Bottom Row:} Mass FP. Despite their structural evolution, galaxies at all redshifts lie on a very stable mass FP.}
	\label{fig:fundplane}
\end{figure*}

\begin{figure*}[!t]
   \centering
	\includegraphics*[width=0.8\textwidth]{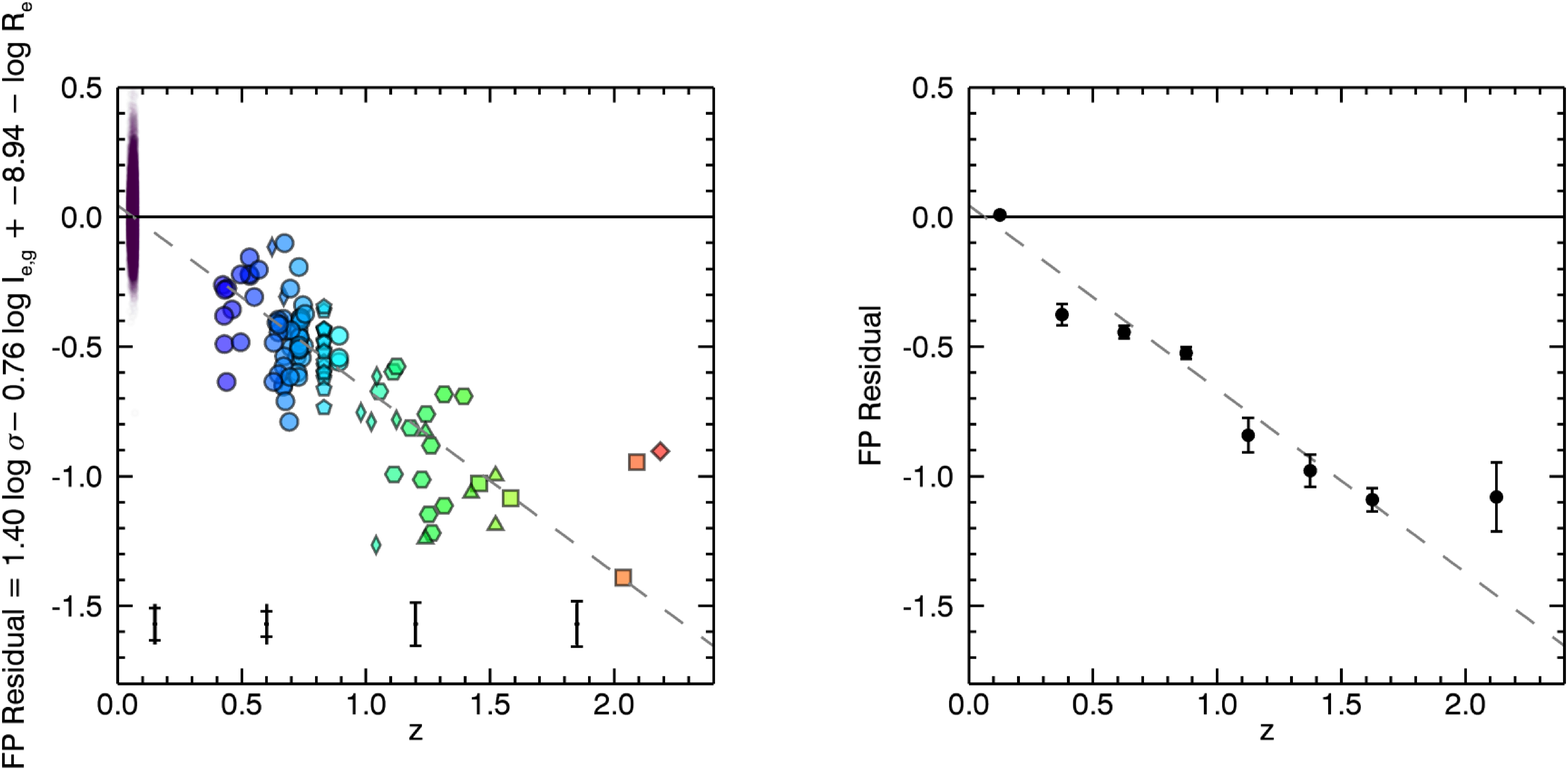}
	\includegraphics*[width=0.8\textwidth]{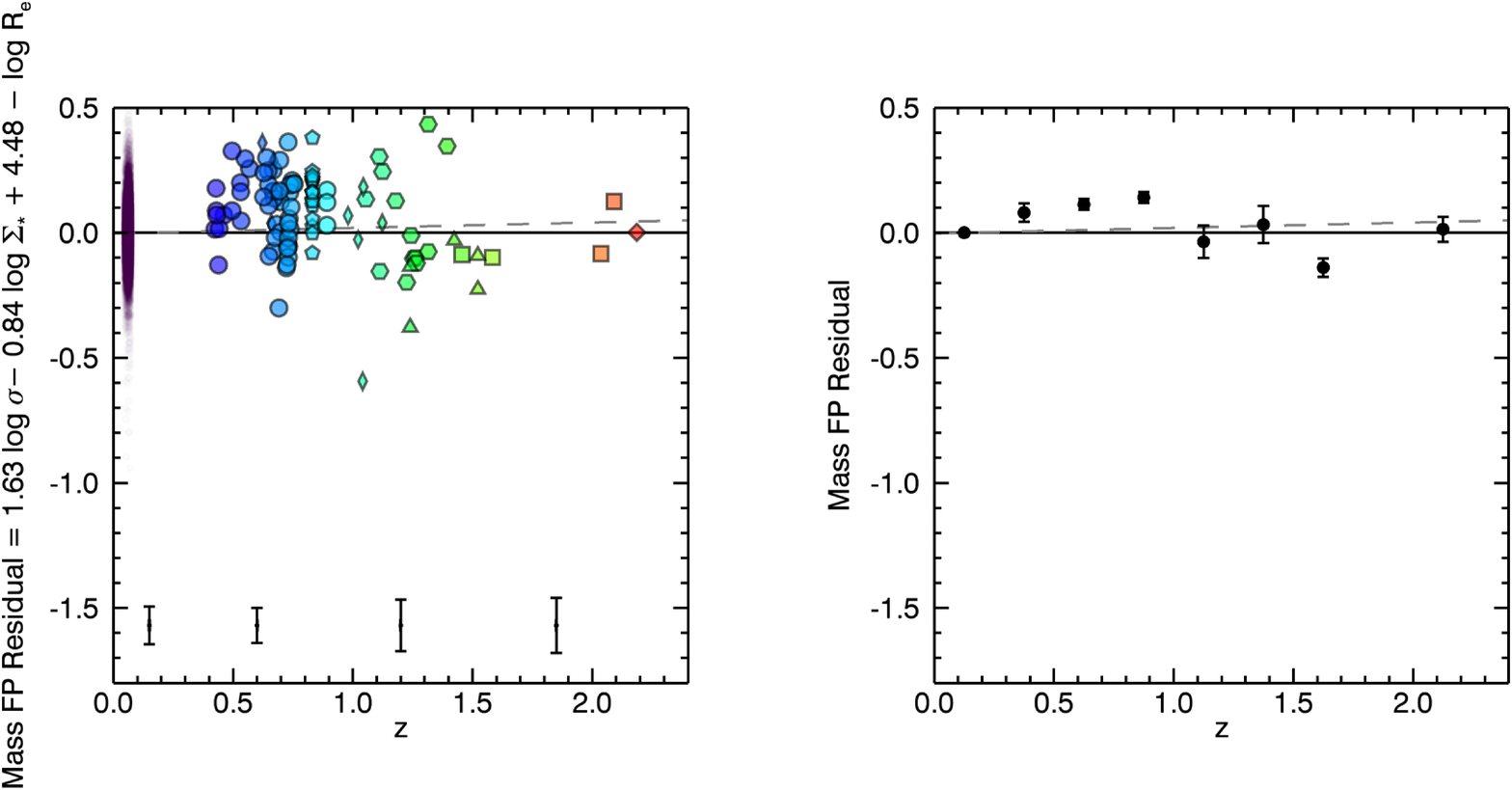}
	\caption{Residuals from the luminosity (top) and mass FP (bottom) with best-fit evolving FPs (dashed gray lines). Left panels include individual galaxies; right  show average residual in redshift bins.  While the FP evolves $\sim1.4\unit{dex}$ since $z\sim2$, mass FP residuals are $\lesssim0.15\unit{dex}$ at all redshifts and evolves only $\sim0.04\unit{dex}$ since $z\sim2$.}
	\label{fig:fp_resid}
\end{figure*}
 
Fig.\ref{fig:fundplane} shows an edge-on view of the $g'$-band FP out to $z\sim2$. The FP follows the form:
\begin{equation}\log\,R_e=a\log\sigma+b\log\,I_{e,g'}+c_{z},\end{equation}
with a zeropoint that scales linearly with redshift: $c_{z}=c+d(z-\langle{z_{SDSS}}\rangle)$ \citep[see e.g.][]{dokkummarel:07} and $\langle{z_{SDSS}}\rangle=0.063$. We define $\log\,I_{e,g'}\equiv-0.4\langle\mu_{e,g'}\rangle$, where $\langle\mu_{e,g'}\rangle$ is the average surface brightness within $R_e$, corrected for cosmological surface brightness dimming (see e.g. HB09) as:
\begin{equation}\langle\mu_{e,g'}\rangle = m_{g'}+2.5\log(2\pi\,{R_e}^2)-10\log(1+z).\end{equation}

Sample incompleteness is significant at high redshift and would bias measurements of the slope of the FP \citep[e.g.][]{jorgensen:96}, however \citet{holden:10} found that the form of the FP does not evolve strongly with redshift. As shown in many other studies, there is strong evolution in the zeropoint \citep[see also e.g.][and references therein]{dokkum:96,dokkummarel:07,holden:10,toft:12}.  We assume the HB09 slope ($[a,b]=[1.404,0.761]$) and fit zeropoint and evolution as follows. First we fit the SDSS zeropoint ($c=c(z=\langle\,z_{SDSS}\rangle)$) by minimizing mean absolute orthogonal residuals from the plane:
\begin{equation}\Delta=\frac{|\log\,R_e-a\log\sigma-b\,I_{e,g'}-c_{z}\,|}{\sqrt{1+a^2+b^2}},\end{equation}
for galaxies in the SDSS sample, with bootstrap resampling to estimate errors \citep[following e.g.][]{jorgensen:96}. For errors on the $z\sim0$ zeropoint, we add $\sim10\%$ error in quadrature to account for systematic sources of error.  Fixing $[a,b,c]$, we draw samples from the SDSS and high-z samples and fit for the redshift evolution.

In the mass FP, we replace $I_{e,g'}$ with stellar mass surface density:
\begin{equation}\Sigma_{\star}\equiv\left(\frac{M_{\star}}{2\pi{R_e}^2}\right)\end{equation}
The bottom row of Fig.\ref{fig:fundplane} shows the evolution of the mass FP, which follows:
\begin{equation}\log\,R_e=\alpha\log\sigma+\beta\log\Sigma_{\star}+\gamma_z,\end{equation} 
in which the redshift-dependent zeropoint scales as $\gamma_z=\gamma+\delta\log{(1+z-\langle{z_{SDSS}\rangle})}$.  Again we assume the HB09 slope $[\alpha,\beta]=[1.6287,-0.840]$ of the mass FP (from $M/L_{spec}$) and orthogonally fit the $z\sim0$ zeropoint $\langle\gamma\rangle=4.475\pm\,0.001$ with bootstrap errors and add $\sim10\%$ systematic error. This fit is similar to the value obtained by HB09 ($\gamma=4.42$), despite subtle differences e.g. in sample selection, aperture corrections, and size measurements.  We fit redshift evolution of the mass FP, $\delta=-0.095\pm0.043$, for the complete sample with bootstrap resampling. 

Best-fit FP and mass FP are included in Fig.\ref{fig:fundplane} and Fig.\ref{fig:fp_resid} (black solid line: $z\sim0$, gray dashed line: redshift-dependent) and reported in Table \ref{tbl:FP}. 

\begin{deluxetable*}{cccccccc}
 	\centering
 	\tablecaption{Best-Fit Fundamental Planes}
	\tabletypesize{\footnotesize}
	\tablewidth{\textwidth}
	\tablehead{\multicolumn{4}{c}{\textbf{Luminosity FP}} & \multicolumn{4}{c}{\textbf{Mass FP}} \\ [+1ex]
			\multicolumn{4}{c}{$\log\!R_e=a\log\sigma+b\,I_{e,g}+c+d(z-0.063)$} & \multicolumn{4}{c}{$\log\!R_e=\alpha\log\sigma+\beta\log\Sigma_{\star}+\gamma+\delta\log{(1+z-0.063)}$} \\ [+1ex]
			\cline{1-4} \cline{5-8} \\ [-1ex]
			\colhead{$a$} & \colhead{$b$} & \colhead{$c$} & \colhead{$d$} & 
			\colhead{$\alpha$} & \colhead{$\beta$} & \colhead{$\gamma$} & \colhead{$\delta$}}
	\startdata
	1.404 & -0.761 & $-8.943\pm0.894$ & $0.708\pm0.016$ & 1.6287 & -0.840 &  $4.475\pm 0.448$ & 
  $-0.095\pm0.043$ \\
	\enddata
	 \label{tbl:FP}
 \end{deluxetable*}

We calculate scatter about $\log\,R_e$ using the biweight sigma to calculate the outlier resistant rms in: 
\begin{equation}
\Delta_{FP}=\log\,R_e-(a\log\sigma+b\,I_{e,g}+c+d(z-0.063))
\end{equation}
and 
\begin{equation}
\Delta_{MFP}=\log\,R_e-(\alpha\log\sigma+\beta\log\Sigma_{\star}+\gamma+\delta\log(1+z-0.063)).
\end{equation}
We measure a scatter of $0.110\pm0.009$ in the SDSS FP and $0.106\pm0.001$ in the mass FP, of which $0.0624$ and $0.076$ respectively are due to observational errors. Uncertainties in observed scatter are estimated with bootstrap resampling and scatter due to measurement errors are averages of simulations of scattered data. We find that scatter increases at $z\gtrsim0$, with $rms_{FP}=0.161\pm0.018$ ($0.066\unit{dex}$ due to errors) and  $rms_{MFP}=0.171\pm0.016$ ($0.086\unit{dex}$ due to errors). Therefore both FPs have similar intrinsic scatter: $0.146\unit{dex}$ about the FP and $0.148\unit{dex}$ about the mass FP.

The evolution of the mass FP is in striking contrast with that of the luminosity FP. The mass FP exhibits very little apparent zeropoint evolution: galaxies out to $z\sim2$ lie roughly along the local relation.  Residuals, calculated by subtracting out the $z\sim0$ luminosity and mass FPs, as a function of redshift are shown in Fig.\ref{fig:fp_resid}, for individual galaxies (left) and averages as a function of redshift (right). By $z\sim2$, the FP zeropoint has evolved by $\sim1.4\unit{dex}$. Residuals from the mass FP do not systematically evolve with redshift; the best-fit relation introduces only $\sim-0.04\unit{dex}$ of evolution by $z\sim2$ and average residuals are within $0.15\unit{dex}$ of the $z\sim0$ relation. 
 
 This result was an implicit assumption in previous studies of FP evolution.  However, the observed strong evolution of massive galaxies since $z\sim2$ in size \citep[e.g.][and references therein]{dokkumnic:08}, morphologies \citep[e.g.][]{wel:11}, and stellar populations might lead one to question this assumption. Despite this complexity, it seems clear that whatever processes dominate the evolution of dynamically massive galaxies in the last $\sim10$ billion years do so within the mass FP.

\section{Minimal Tilt of the Mass FP}

\citet{faber:87} first demonstrated that the existence of the FP can be understood as a consequence of the Virial theorem, plus a tilt. The stability of the mass FP presented in this Letter is evidence that offsets of the classical FP relative to $z\sim0$ are largely due to $M/L$ variations. However, whether the mass FP is consistent with the Virial Plane, $R_e\propto\,\sigma^2\,\Sigma_e^{-1}$,  \citep[e.g.][]{cappellari:12} or slightly tilted due to other effects such as non-homology (HB09) is debated.

\begin{figure*}[!t]
\centering
	\includegraphics*[width=0.24\textwidth]{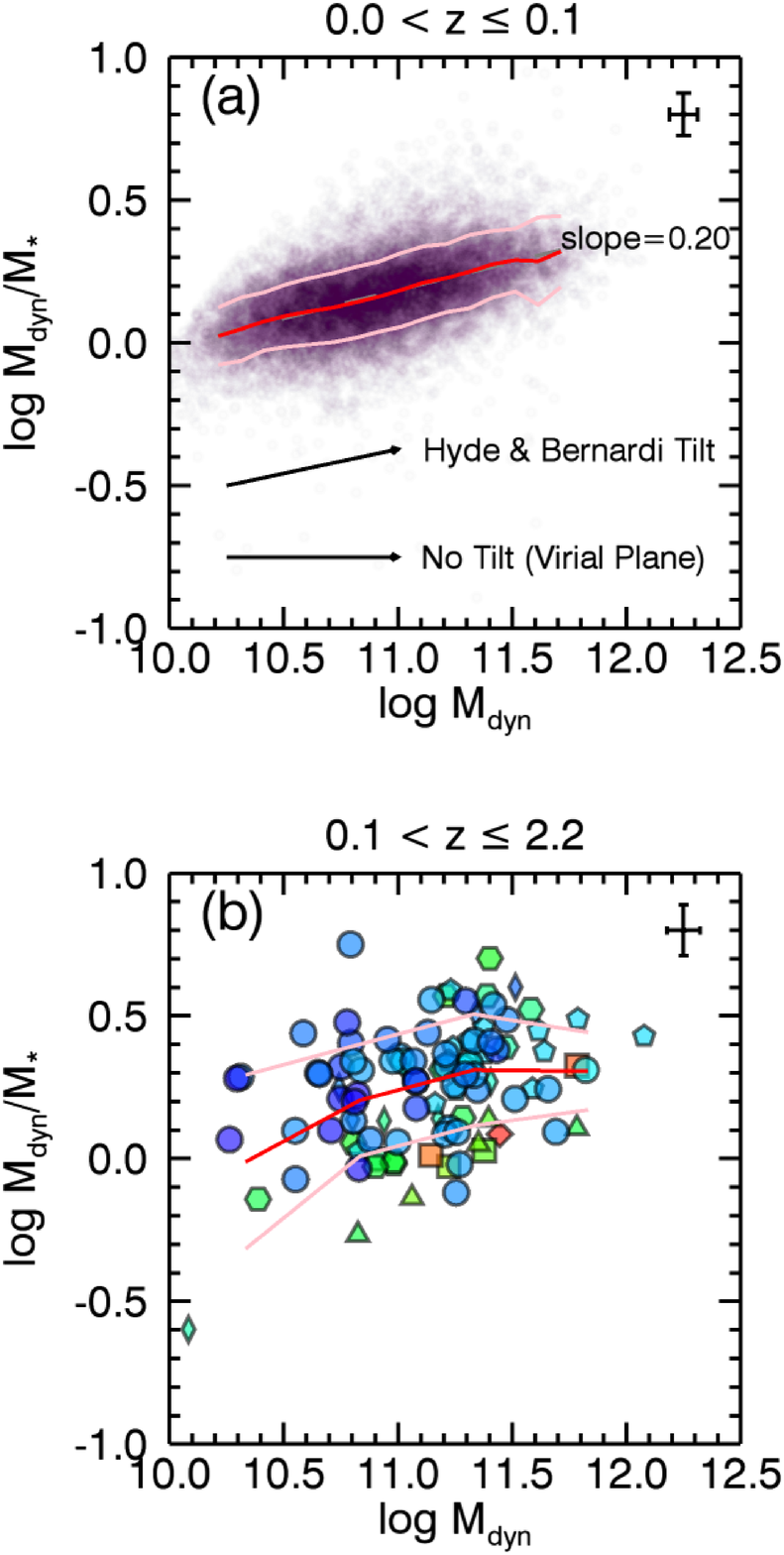}
	\includegraphics*[width=0.24\textwidth]{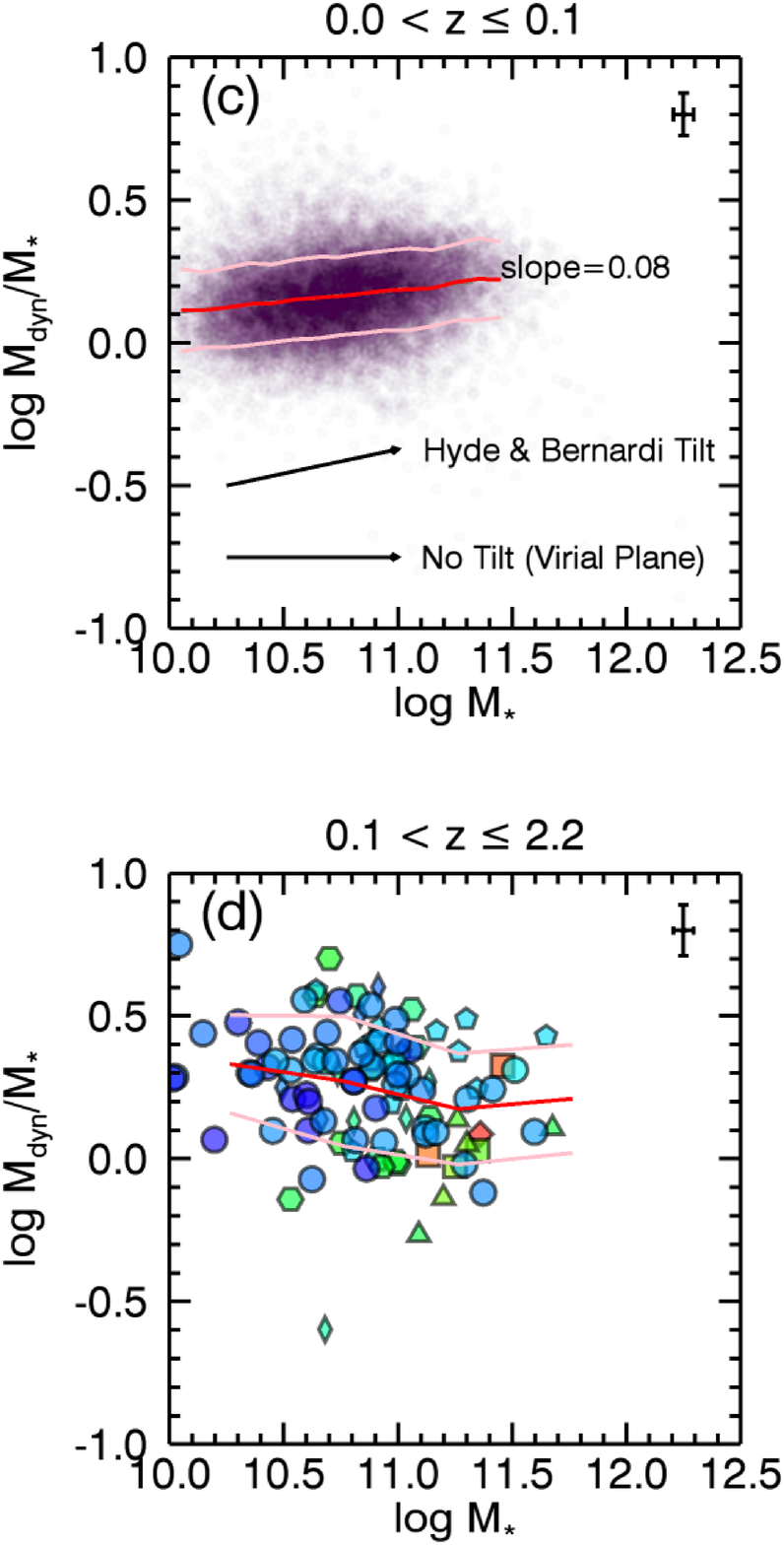}
	\includegraphics*[width=0.24\textwidth]{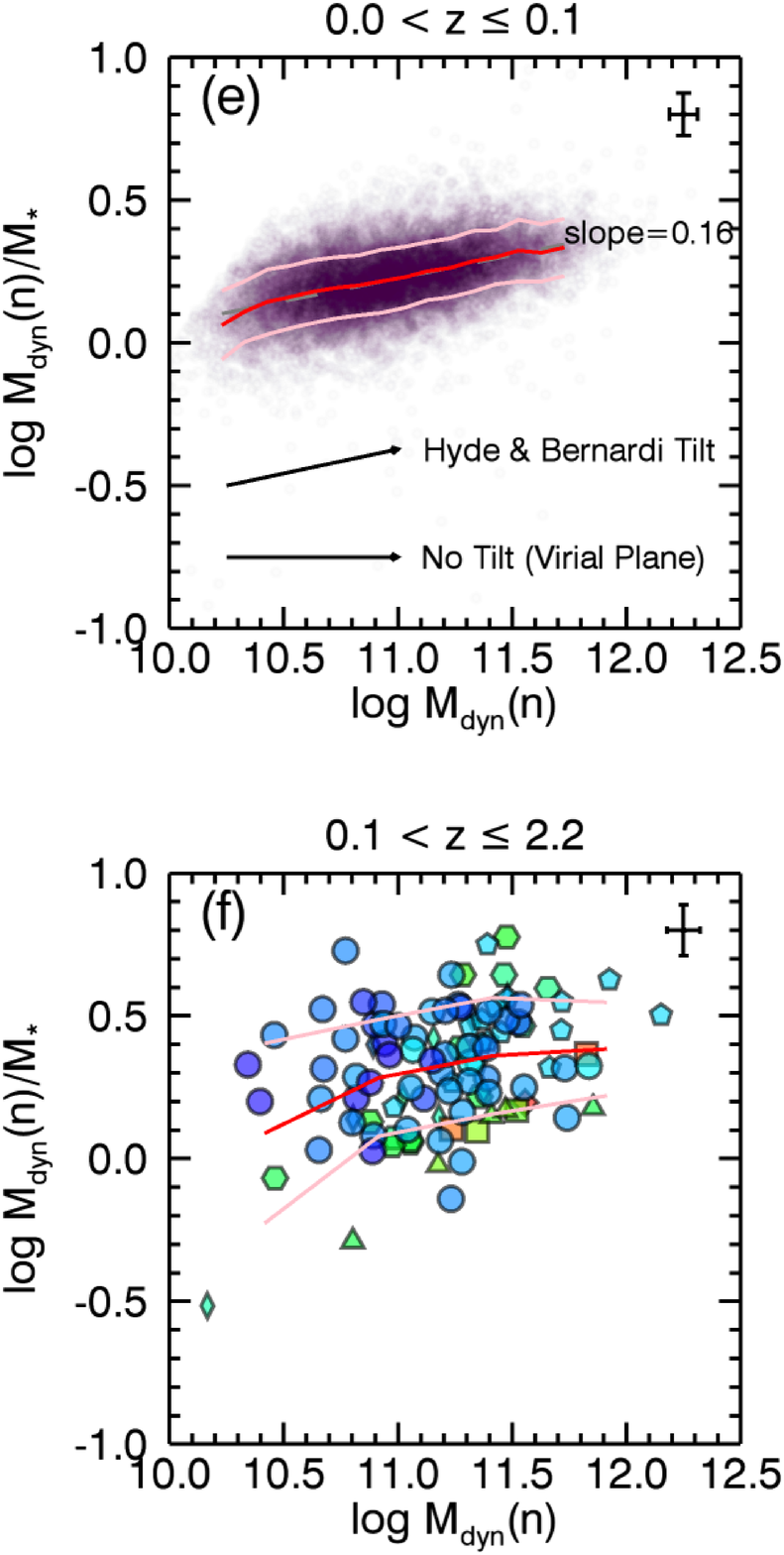}
	\includegraphics*[width=0.24\textwidth]{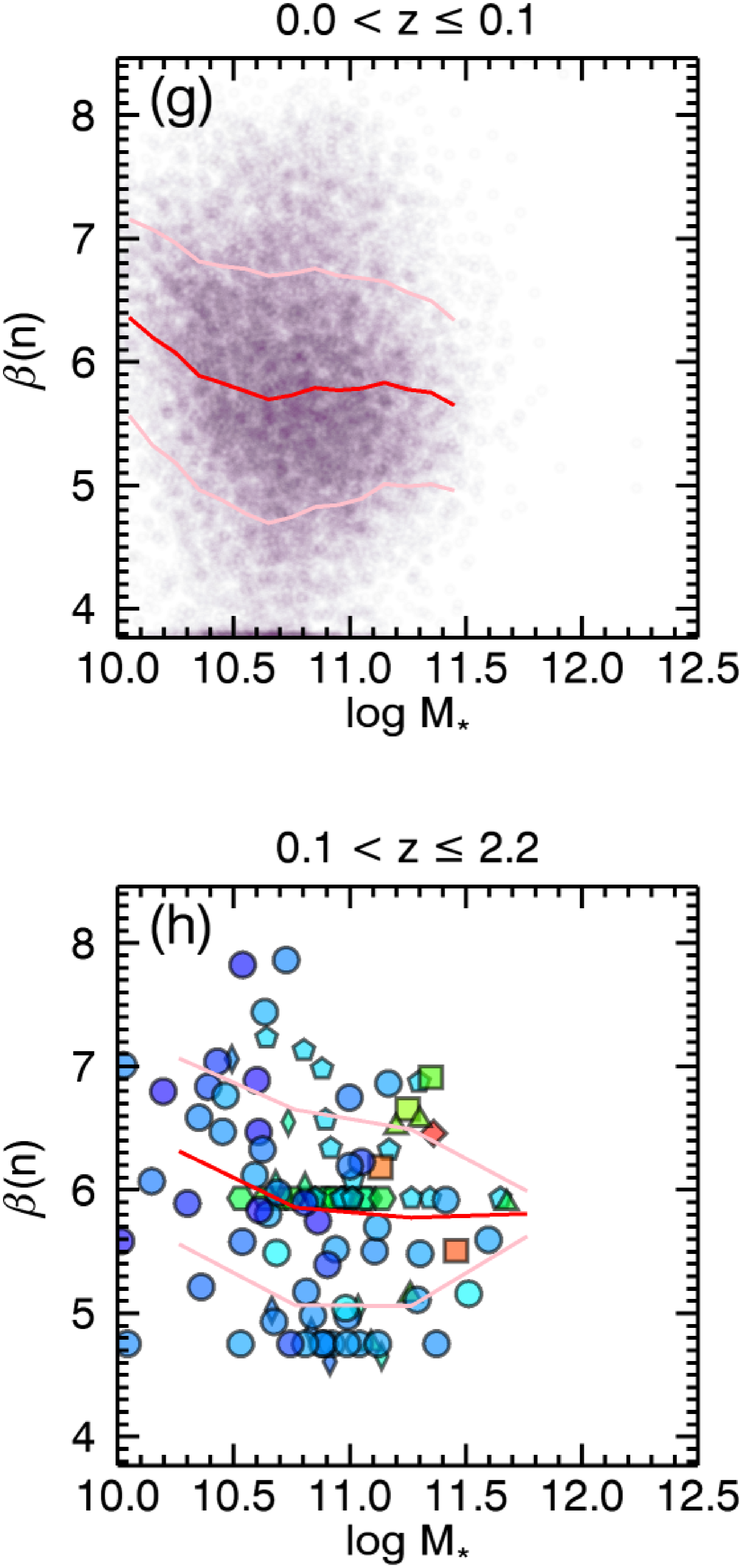}
	\label{fig:mfp_tilt}
	\caption{$M_{dyn}/M_{\star}$ versus mass to assess mass FP tilt. (a-b) $M_{dyn}/M_{\star}$ versus $M_{dyn}$ as in HB09, (c-d) $M_{dyn}/M_{\star}$ versus $M_{\star}$, and (e-f) $M_{dyn}(n)/M_{\star}$ versus $M_{dyn}(n)$, adopting a S\'ersic-dependent virial constant (g-h). A slight tilt in the mass FP is observed when compared to $M_{dyn}$, but this tilt is not obvious in all projections and we emphasize the need for full modeling of selection biases.}
\end{figure*}

Fig.\ref{fig:mfp_tilt} shows $M_{dyn}/M_{\star}$ versus mass to assess the mass FP tilt. In this case a tilted mass FP would result from variations of $M_{dyn}/M_{\star}$ within the galaxy population. In Fig.\ref{fig:mfp_tilt}\,a-b, we reproduce the HB09 tilt, plotting against $\log\,M_{dyn}=5\sigma^2\,R_e/G$.  However, when compared to $M_{\star}$ (Fig.\ref{fig:mfp_tilt}\,c-d), the tilt disappears and the relation is approximately Virial. In Fig.\ref{fig:mfp_tilt}\,e-f, we introduce a S\'ersic-dependent Virial constant, $\beta(n)=8.87-0.831n+0.0241n^2$ \citep{cappellari:06} (Fig.\ref{fig:mfp_tilt}\,g-h), to determine $M_{dyn}(n)$, again finding a slight tilt. We note that the distribution of $\beta(n)$ in $M_{\star}$ appears to be similar for the high and low-z galaxies in this sample. \citet{taylor:10} found that adopting a S\'ersic-dependent Virial constant (for $\sigma(<R_e/8)$) largely removes the $M_{dyn}/M_{\star}$ variations with \emph{stellar} mass. We do not find strong variation in $M_{dyn}/M_{\star}$ across the galaxy populations (for $\sigma(<R_e)$). However, since introducing $\beta(n)$ results in a slightly shallower tilt ($M_{dyn}(n)/M_{\star}\propto\,M_{dyn}(n)^{0.16}$ (Fig.\ref{fig:mfp_tilt}\,e) versus $M_{dyn}/M_{\star}\propto\,M_{dyn}^{0.20}$ (Fig.\ref{fig:mfp_tilt}\,a) in SDSS) non-homology may still be important to the mass FP tilt. 
We emphasize that sample incompleteness can influence these diagnostics dramatically; for example a lack of low $M_{\star}$ galaxies could introduce the dearth of high $M_{dyn}/M_{\star}$ ratios at low $M_{dyn}$. Sample cuts in magnitude, signal-to-noise, S\'ersic index, and velocity dispersion will also influence slope measurements. We conclude that there is no strong evidence for a tilt of the mass FP in this sample and emphasize the need for more complete spectroscopic samples at high-z and a full model including selection effects and distribution functions to resolve this issue.

\section{Discussion and Conclusions}

In this Letter, we find that the mass FP, or relation between stellar mass density, size, and velocity dispersion, for massive galaxies has not evolved significantly since $z\sim2$. This is in striking contrast with the evolution of the luminosity FP, which was also in place but with an offset. This result is noteworthy because of the observed strong structural evolution of massive galaxies in this epoch.

The mass FP reflects the combination of the virial theorem, the relation between $M_{dyn}/M_{\star}$ and $M_{dyn}$ (i.e. the dark matter fraction as a function of $M_{dyn}$), and the non-homology as a function of $M_{dyn}$. \citet{robertson:06a} suggest that the tilt in the mass FP may arise from quenching via gas-rich mergers: gas cooling and star formation are more efficient in low mass systems, leading to an increase of $M_{dyn}/M_{\star}$ with increasing $M_{dyn}$. \citet{robertson:06a} also find that gas-poor mergers do not alter the tilt of the mass FP \citep[see also e.g.][]{boylan:06}. These results are in qualitative agreement with the observational results presented in this Letter. In this context, the weak or absent trends with redshift in Fig. \ref{fig:mfp_tilt} may suggest that dissipationless processes dominate in our samples.

This study ignores the potential impact of non-homology on the mass FP. Evolution in S\'ersic index and effective radius can have significant implications for evolving dark matter fractions in massive galaxies. Although this will be unimportant for measured size and surface density, velocity dispersion is sensitive to the total mass enclosed within a given radius.  Given the observed inside-out growth of massive galaxies \citep[][]{bezanson:09,patel:13a} the impact of this growth on the mass FP could be dramatic. However, the magnitude of this effect is strongly dependent on the evolution of the stellar and dark matter profiles, the latter being extremely difficult to constrain. Models of galaxy evolution including observational constraints on the structural evolution of massive galaxies as well as realistic dark matter halos would be extremely informative with respect to the impact of non-homology on the mass FP.

A number of observational studies have found evidence for non-universal IMFs \citep[e.g.][]{dokkumconroy:10}, indicating that dwarf populations, and therefore M/Ls, increase with the depth of galaxy potential wells. The assumption that galaxies follow a \citet{chabrier:03} IMF could further contribute to the apparent tilt in $M_{\star}/M_{dyn}$ versus $M_{dyn}$. 

Finally, we emphasize that this high-redshift sample is quite biased towards the most massive, brightest galaxies. Larger, more representative spectroscopic samples of galaxy dynamics at $z>1.5$ are becoming possible with the newest generation of multi-object, near-IR spectrographs. Future studies will help determine the scatter in galaxy properties along the mass FP at lower masses and out to higher redshifts.

\begin{acknowledgements}

The authors wish to thank Frank van den Bosch, Aaron Dutton, and Ravi Sheth for interesting discussions which contributed to this Letter and Andrew Newman for sharing ancillary data for his spectroscopic sample. The authors are grateful for access to the Keck telescopes on Mauna Kea and to all contributing agencies for funding the SDSS survey. The analysis pipeline used to reduce the DEIMOS data was developed at UC Berkeley with support from NSF grant AST-0071048.

\end{acknowledgements}


\begin{thebibliography}{}
\expandafter\ifx\csname natexlab\endcsname\relax\def\natexlab#1{#1}\fi

\bibitem[{{Abazajian} {et~al.}(2009){Abazajian}, {Adelman-McCarthy},
  {Ag{\"u}eros}, {Allam}, {Allende Prieto}, {An}, {Anderson}, {Anderson},
  {Annis}, {Bahcall}, \& et~al.}]{dr7}
{Abazajian}, K.~N., {Adelman-McCarthy}, J.~K., {Ag{\"u}eros}, M.~A., {et~al.}
  2009, \apjs, 182, 543

\bibitem[{{Bezanson} {et~al.}(2013){Bezanson}, {van Dokkum}, {van de Sande},
  {Franx}, \& {Kriek}}]{bezanson:13}
{Bezanson}, R., {van Dokkum}, P., {van de Sande}, J., {Franx}, M., \& {Kriek},
  M. 2013, \apjl, 764, L8

\bibitem[{{Bezanson} {et~al.}(2009){Bezanson}, {van Dokkum}, {Tal},
  {Marchesini}, {Kriek}, {Franx}, \& {Coppi}}]{bezanson:09}
{Bezanson}, R., {van Dokkum}, P.~G., {Tal}, T., {et~al.} 2009, \apj, 697, 1290

\bibitem[{{Bezanson} {et~al.}(2011){Bezanson}, {van Dokkum}, {Franx},
  {Brammer}, {Brinchmann}, {Kriek}, {Labb{\'e}}, {Quadri}, {Rix}, {van de
  Sande}, {Whitaker}, \& {Williams}}]{bezanson:11}
{Bezanson}, R., {van Dokkum}, P.~G., {Franx}, M., {et~al.} 2011, \apjl, 737,
  L31+

\bibitem[{{Blakeslee} {et~al.}(2006){Blakeslee}, {Holden}, {Franx}, {Rosati},
  {Bouwens}, {Demarco}, {Ford}, {Homeier}, {Illingworth}, {Jee}, {Mei},
  {Menanteau}, {Meurer}, {Postman}, \& {Tran}}]{blakeslee:06}
{Blakeslee}, J.~P., {Holden}, B.~P., {Franx}, M., {et~al.} 2006, \apj, 644, 30

\bibitem[{{Blanc} {et~al.}(2008){Blanc}, {Lira}, {Barrientos}, {Aguirre},
  {Francke}, {Taylor}, {Quadri}, {Marchesini}, {Infante}, {Gawiser}, {Hall},
  {Willis}, {Herrera}, \& {Maza}}]{blanc:08}
{Blanc}, G.~A., {Lira}, P., {Barrientos}, L.~F., {et~al.} 2008, \apj, 681, 1099

\bibitem[{{Blanton} \& {Roweis}(2007)}]{blanton:07}
{Blanton}, M.~R., \& {Roweis}, S. 2007, \aj, 133, 734

\bibitem[{{Boylan-Kolchin} {et~al.}(2006){Boylan-Kolchin}, {Ma}, \&
  {Quataert}}]{boylan:06}
{Boylan-Kolchin}, M., {Ma}, C.-P., \& {Quataert}, E. 2006, \mnras, 369, 1081

\bibitem[{{Brinchmann} {et~al.}(2004){Brinchmann}, {Charlot}, {White},
  {Tremonti}, {Kauffmann}, {Heckman}, \& {Brinkmann}}]{brinchmann:04}
{Brinchmann}, J., {Charlot}, S., {White}, S.~D.~M., {et~al.} 2004, \mnras, 351,
  1151

\bibitem[{{Bruzual} \& {Charlot}(2003)}]{bc:03}
{Bruzual}, G., \& {Charlot}, S. 2003, \mnras, 344, 1000

\bibitem[{{Cappellari} \& {Emsellem}(2004)}]{cappellari:04}
{Cappellari}, M., \& {Emsellem}, E. 2004, \pasp, 116, 138

\bibitem[{{Cappellari} {et~al.}(2006){Cappellari}, {Bacon}, {Bureau}, {Damen},
  {Davies}, {de Zeeuw}, {Emsellem}, {Falc{\'o}n-Barroso}, {Krajnovi{\'c}},
  {Kuntschner}, {McDermid}, {Peletier}, {Sarzi}, {van den Bosch}, \& {van de
  Ven}}]{cappellari:06}
{Cappellari}, M., {Bacon}, R., {Bureau}, M., {et~al.} 2006, \mnras, 366, 1126

\bibitem[{{Cappellari} {et~al.}(2012){Cappellari}, {McDermid}, {Alatalo},
  {Blitz}, {Bois}, {Bournaud}, {Bureau}, {Crocker}, {Davies}, {Davis}, {de
  Zeeuw}, {Duc}, {Emsellem}, {Khochfar}, {Krajnovi{\'c}}, {Kuntschner},
  {Lablanche}, {Morganti}, {Naab}, {Oosterloo}, {Sarzi}, {Scott}, {Serra},
  {Weijmans}, \& {Young}}]{cappellari:12}
{Cappellari}, M., {McDermid}, R.~M., {Alatalo}, K., {et~al.} 2012, \nat, 484,
  485

\bibitem[{{Chabrier}(2003)}]{chabrier:03}
{Chabrier}, G. 2003, \pasp, 115, 763

\bibitem[{{Chevance} {et~al.}(2012){Chevance}, {Weijmans}, {Damjanov},
  {Abraham}, {Simard}, {van den Bergh}, {Caris}, \& {Glazebrook}}]{chevance:12}
{Chevance}, M., {Weijmans}, A.-M., {Damjanov}, I., {et~al.} 2012, \apjl, 754,
  L24

\bibitem[{{Cooper} {et~al.}(2012){Cooper}, {Newman}, {Davis}, {Finkbeiner}, \&
  {Gerke}}]{cooper:12}
{Cooper}, M.~C., {Newman}, J.~A., {Davis}, M., {Finkbeiner}, D.~P., \& {Gerke},
  B.~F. 2012, {spec2d: DEEP2 DEIMOS Spectral Pipeline}, astrophysics Source
  Code Library, ascl:1203.003

\bibitem[{{Djorgovski} \& {Davis}(1987)}]{djorgovski:87}
{Djorgovski}, S., \& {Davis}, M. 1987, \apj, 313, 59

\bibitem[{{Dressler} {et~al.}(1987){Dressler}, {Lynden-Bell}, {Burstein},
  {Davies}, {Faber}, {Terlevich}, \& {Wegner}}]{dressler:87}
{Dressler}, A., {Lynden-Bell}, D., {Burstein}, D., {et~al.} 1987, \apj, 313, 42

\bibitem[{{Faber}(1987)}]{faber:87}
{Faber}, S.~M., ed. 1987, {Nearly normal galaxies: From the Planck time to the
  present; Proceedings of the Eighth Santa Cruz Summer Workshop in Astronomy
  and Astrophysics, Santa Cruz, CA, July 21-Aug. 1, 1986}

\bibitem[{{F{\"o}rster Schreiber} {et~al.}(2006){F{\"o}rster Schreiber},
  {Franx}, {Labb{\'e}}, {Rudnick}, {van Dokkum}, {Illingworth}, {Kuijken},
  {Moorwood}, {Rix}, {R{\"o}ttgering}, \& {van der Werf}}]{forster:06}
{F{\"o}rster Schreiber}, N.~M., {Franx}, M., {Labb{\'e}}, I., {et~al.} 2006,
  \aj, 131, 1891

\bibitem[{{Graves} {et~al.}(2007){Graves}, {Faber}, {Schiavon}, \&
  {Yan}}]{graves:07}
{Graves}, G.~J., {Faber}, S.~M., {Schiavon}, R.~P., \& {Yan}, R. 2007, \apj,
  671, 243

\bibitem[{{Grogin} {et~al.}(2011){Grogin}, {Kocevski}, {Faber}, {Ferguson},
  {Koekemoer}, {Riess}, {Acquaviva}, {Alexander}, {Almaini}, {Ashby}, {Barden},
  {Bell}, {Bournaud}, {Brown}, {Caputi}, {Casertano}, {Cassata}, {Castellano},
  {Challis}, {Chary}, {Cheung}, {Cirasuolo}, {Conselice}, {Roshan Cooray},
  {Croton}, {Daddi}, {Dahlen}, {Dav{\'e}}, {de Mello}, {Dekel}, {Dickinson},
  {Dolch}, {Donley}, {Dunlop}, {Dutton}, {Elbaz}, {Fazio}, {Filippenko},
  {Finkelstein}, {Fontana}, {Gardner}, {Garnavich}, {Gawiser}, {Giavalisco},
  {Grazian}, {Guo}, {Hathi}, {H{\"a}ussler}, {Hopkins}, {Huang}, {Huang},
  {Jha}, {Kartaltepe}, {Kirshner}, {Koo}, {Lai}, {Lee}, {Li}, {Lotz}, {Lucas},
  {Madau}, {McCarthy}, {McGrath}, {McIntosh}, {McLure}, {Mobasher},
  {Moustakas}, {Mozena}, {Nandra}, {Newman}, {Niemi}, {Noeske}, {Papovich},
  {Pentericci}, {Pope}, {Primack}, {Rajan}, {Ravindranath}, {Reddy}, {Renzini},
  {Rix}, {Robaina}, {Rodney}, {Rosario}, {Rosati}, {Salimbeni}, {Scarlata},
  {Siana}, {Simard}, {Smidt}, {Somerville}, {Spinrad}, {Straughn}, {Strolger},
  {Telford}, {Teplitz}, {Trump}, {van der Wel}, {Villforth}, {Wechsler},
  {Weiner}, {Wiklind}, {Wild}, {Wilson}, {Wuyts}, {Yan}, \& {Yun}}]{candels}
{Grogin}, N.~A., {Kocevski}, D.~D., {Faber}, S.~M., {et~al.} 2011, \apjs, 197,
  35

\bibitem[{{Holden} {et~al.}(2010){Holden}, {van der Wel}, {Kelson}, {Franx}, \&
  {Illingworth}}]{holden:10}
{Holden}, B.~P., {van der Wel}, A., {Kelson}, D.~D., {Franx}, M., \&
  {Illingworth}, G.~D. 2010, \apj, 724, 714

\bibitem[{{Hyde} \& {Bernardi}(2009)}]{hyde:09}
{Hyde}, J.~B., \& {Bernardi}, M. 2009, \mnras, 396, 1171

\bibitem[{{Jorgensen} {et~al.}(1996){Jorgensen}, {Franx}, \&
  {Kjaergaard}}]{jorgensen:96}
{Jorgensen}, I., {Franx}, M., \& {Kjaergaard}, P. 1996, \mnras, 280, 167

\bibitem[{{Kelson} {et~al.}(2000){Kelson}, {Illingworth}, {van Dokkum}, \&
  {Franx}}]{kelsonfp:00}
{Kelson}, D.~D., {Illingworth}, G.~D., {van Dokkum}, P.~G., \& {Franx}, M.
  2000, \apj, 531, 184

\bibitem[{{Kriek} {et~al.}(2009){Kriek}, {van Dokkum}, {Labb{\'e}}, {Franx},
  {Illingworth}, {Marchesini}, \& {Quadri}}]{kriek:09}
{Kriek}, M., {van Dokkum}, P.~G., {Labb{\'e}}, I., {et~al.} 2009, \apj, 700,
  221

\bibitem[{{Newman} {et~al.}(2010){Newman}, {Ellis}, {Treu}, \&
  {Bundy}}]{newman:10}
{Newman}, A.~B., {Ellis}, R.~S., {Treu}, T., \& {Bundy}, K. 2010, \apjl, 717,
  L103

\bibitem[{{Newman} {et~al.}(2012){Newman}, {Cooper}, {Davis}, {Faber}, {Coil},
  {Guhathakurta}, {Koo}, {Phillips}, {Conroy}, {Dutton}, {Finkbeiner}, {Gerke},
  {Rosario}, {Weiner}, {Willmer}, {Yan}, {Harker}, {Kassin}, {Konidaris},
  {Lai}, {Madgwick}, {Noeske}, {Wirth}, {Connolly}, {Kaiser}, {Kirby},
  {Lemaux}, {Lin}, {Lotz}, {Luppino}, {Marinoni}, {Matthews}, {Metevier}, \&
  {Schiavon}}]{newmandeep:12}
{Newman}, J.~A., {Cooper}, M.~C., {Davis}, M., {et~al.} 2012, arXiv:1203.3192,
  arXiv:1203.3192

\bibitem[{{Patel} {et~al.}(2013){Patel}, {van Dokkum}, {Franx}, {Quadri},
  {Muzzin}, {Marchesini}, {Williams}, {Holden}, \& {Stefanon}}]{patel:13a}
{Patel}, S.~G., {van Dokkum}, P.~G., {Franx}, M., {et~al.} 2013, \apj, 766, 15

\bibitem[{{Robertson} {et~al.}(2006){Robertson}, {Cox}, {Hernquist}, {Franx},
  {Hopkins}, {Martini}, \& {Springel}}]{robertson:06a}
{Robertson}, B., {Cox}, T.~J., {Hernquist}, L., {et~al.} 2006, \apj, 641, 21

\bibitem[{{Saglia} {et~al.}(2010){Saglia}, {S{\'a}nchez-Bl{\'a}zquez},
  {Bender}, {Simard}, {Desai}, {Arag{\'o}n-Salamanca}, {Milvang-Jensen},
  {Halliday}, {Jablonka}, {Noll}, {Poggianti}, {Clowe}, {De Lucia},
  {Pell{\'o}}, {Rudnick}, {Valentinuzzi}, {White}, \& {Zaritsky}}]{saglia:10}
{Saglia}, R.~P., {S{\'a}nchez-Bl{\'a}zquez}, P., {Bender}, R., {et~al.} 2010,
  \aap, 524, A6

\bibitem[{{Simard} {et~al.}(2011){Simard}, {Mendel}, {Patton}, {Ellison}, \&
  {McConnachie}}]{simard:11}
{Simard}, L., {Mendel}, J.~T., {Patton}, D.~R., {Ellison}, S.~L., \&
  {McConnachie}, A.~W. 2011, \apjs, 196, 11

\bibitem[{{Taylor} {et~al.}(2010){Taylor}, {Franx}, {Brinchmann}, {van der
  Wel}, \& {van Dokkum}}]{taylor:10}
{Taylor}, E.~N., {Franx}, M., {Brinchmann}, J., {van der Wel}, A., \& {van
  Dokkum}, P.~G. 2010, \apj, 722, 1

\bibitem[{{Taylor} {et~al.}(2009){Taylor}, {Franx}, {van Dokkum}, {Quadri},
  {Gawiser}, {Bell}, {Barrientos}, {Blanc}, {Castander}, {Damen},
  {Gonzalez-Perez}, {Hall}, {Herrera}, {Hildebrandt}, {Kriek}, {Labb{\'e}},
  {Lira}, {Maza}, {Rudnick}, {Treister}, {Urry}, {Willis}, \&
  {Wuyts}}]{taylor:09}
{Taylor}, E.~N., {Franx}, M., {van Dokkum}, P.~G., {et~al.} 2009, \apjs, 183,
  295

\bibitem[{{Toft} {et~al.}(2012){Toft}, {Gallazzi}, {Zirm}, {Wold}, {Zibetti},
  {Grillo}, \& {Man}}]{toft:12}
{Toft}, S., {Gallazzi}, A., {Zirm}, A., {et~al.} 2012, \apj, 754, 3

\bibitem[{{van de Sande} {et~al.}(2011){van de Sande}, {Kriek}, {Franx}, {van
  Dokkum}, {Bezanson}, {Whitaker}, {Brammer}, {Labb{\'e}}, {Groot}, \&
  {Kaper}}]{sande:11}
{van de Sande}, J., {Kriek}, M., {Franx}, M., {et~al.} 2011, \apjl, 736, L9

\bibitem[{{van de Sande} {et~al.}(2013){van de Sande}, {Kriek}, {Franx}, {van
  Dokkum}, {Bezanson}, {Bouwens}, {Quadri}, {Rix}, \& {Skelton}}]{sande:13}
---. 2013, \apj, 771, 85

\bibitem[{{van der Wel} {et~al.}(2005){van der Wel}, {Franx}, {van Dokkum},
  {Rix}, {Illingworth}, \& {Rosati}}]{wel:05}
{van der Wel}, A., {Franx}, M., {van Dokkum}, P.~G., {et~al.} 2005, \apj, 631,
  145

\bibitem[{{van der Wel} {et~al.}(2011){van der Wel}, {Rix}, {Wuyts}, {McGrath},
  {Koekemoer}, {Bell}, {Holden}, {Robaina}, \& {McIntosh}}]{wel:11}
{van der Wel}, A., {Rix}, H.-W., {Wuyts}, S., {et~al.} 2011, \apj, 730, 38

\bibitem[{{van der Wel} {et~al.}(2012){van der Wel}, {Bell}, {H{\"a}ussler},
  {McGrath}, {Chang}, {Guo}, {McIntosh}, {Rix}, {Barden}, {Cheung}, {Faber},
  {Ferguson}, {Galametz}, {Grogin}, {Hartley}, {Kartaltepe}, {Kocevski},
  {Koekemoer}, {Lotz}, {Mozena}, {Peth}, \& {Peng}}]{wel:12}
{van der Wel}, A., {Bell}, E.~F., {H{\"a}ussler}, B., {et~al.} 2012, \apjs,
  203, 24

\bibitem[{{van Dokkum} \& {Conroy}(2010)}]{dokkumconroy:10}
{van Dokkum}, P.~G., \& {Conroy}, C. 2010, \nat, 468, 940

\bibitem[{{van Dokkum} \& {Franx}(1996)}]{dokkum:96}
{van Dokkum}, P.~G., \& {Franx}, M. 1996, \mnras, 281, 985

\bibitem[{{van Dokkum} {et~al.}(2009){van Dokkum}, {Kriek}, \&
  {Franx}}]{dokkumnature:09}
{van Dokkum}, P.~G., {Kriek}, M., \& {Franx}, M. 2009, \nat, 460, 717

\bibitem[{{van Dokkum} \& {van der Marel}(2007)}]{dokkummarel:07}
{van Dokkum}, P.~G., \& {van der Marel}, R.~P. 2007, \apj, 655, 30

\bibitem[{{van Dokkum} {et~al.}(2008){van Dokkum}, {Franx}, {Kriek}, {Holden},
  {Illingworth}, {Magee}, {Bouwens}, {Marchesini}, {Quadri}, {Rudnick},
  {Taylor}, \& {Toft}}]{dokkumnic:08}
{van Dokkum}, P.~G., {Franx}, M., {Kriek}, M., {et~al.} 2008, \apjl, 677, L5

\bibitem[{{Whitaker} {et~al.}(2012){Whitaker}, {Kriek}, {van Dokkum},
  {Bezanson}, {Brammer}, {Franx}, \& {Labb{\'e}}}]{whitaker:12a}
{Whitaker}, K.~E., {Kriek}, M., {van Dokkum}, P.~G., {et~al.} 2012, \apj, 745,
  179

\bibitem[{Whitaker {et~al.}(2011)Whitaker, Labbe, van Dokkum, Brammer, Kriek,
  Marchesini, Quadri, Franx, Muzzin, Williams, Bezanson, Illingworth, Lee,
  Lundgren, Nelson, Rudnick, Tal, \& Wake}]{whitaker:11}
Whitaker, K.~E., Labbe, I., van Dokkum, P.~G., {et~al.} 2011, \apj, 735, 86

\bibitem[{{Williams} {et~al.}(2009){Williams}, {Quadri}, {Franx}, {van Dokkum},
  \& {Labb{\'e}}}]{williams:09}
{Williams}, R.~J., {Quadri}, R.~F., {Franx}, M., {van Dokkum}, P., \&
  {Labb{\'e}}, I. 2009, \apj, 691, 1879

\bibitem[{{Wuyts} {et~al.}(2004){Wuyts}, {van Dokkum}, {Kelson}, {Franx}, \&
  {Illingworth}}]{wuyts:04}
{Wuyts}, S., {van Dokkum}, P.~G., {Kelson}, D.~D., {Franx}, M., \&
  {Illingworth}, G.~D. 2004, \apj, 605, 677

\end{thebibliography}
\end{document}